# A NEW STREAM CIPHER:   DICING

Li An-Ping

Beijing 100085, P.R.China
apli0001@sina.com

**Abstract:** In this paper, we will propose a new synchronous stream cipher named DICING, which can be viewed as a clock-controlled one but with a new mechanism of altering steps. With the simple construction, DICING has satisfactory performance, faster than AES about two times. For the security, there have not been found weakness for the known attacks, the key sizes can be 128bits and 256bits respectively.
.

**Keywords:**   stream cipher, LFSR, projector, finite field, correlation attack, algebraic attack, distinguishing attack.

## 1. Introduction

In a synchronous stream cipher, the ciphertext is generally made by bitwise adding (XOR) the plaintext with a binary sequence called keystream. In case that the cipher is abused or the plaintext of some ciphertext are known by some people, and so the keystream will become visible for them, the analysis for this case is called the plaintext-known analysis, a secure keystream should satisfy two basic conditions: The first is that the original key can not be recovered from the keystream, the second is that the contents of the bits in the keystream should be unpredictable for an adversary, in other words, for the adversaries the keystream should look like a random one, i.e. pseudo-random. Clearly, if the keystream sequence is periodic and the period is short, then it will be predictable, thus the keystream should have enough large period. It is known that the technique of the linear feedback shift registers (LFSR) is able to generate the larger periods of sequences, so LFSRs are often used in the stream ciphers as the component parts. However, LFSR also has an obvious weakness that the each bit in a LFSR's sequence is linearly related to the initial state, and so this implies that the initial state is easy deduced from some of the later bits in the sequence, the famous Berlekamp-Massey's algorithm is such a example of the algorithms. In the almost of known attacks such as correlation attacks, algebraic attacks and distinguishing attacks, etc. just exploited the weakness of LFSR. So, LFSR-based stream ciphers should interfere the linear relations in the bits of the LFSRs, the clock-controlled methods comes from this consideration.

The proposal cipher DICING may be taken as a clock-controlled one, but with a new mechanism of altering steps. It consists of a controller and a combiner. In the proposal cipher, we will substitute the LFSR with the LFSR-*like* called projector (Pr.). A projector consists of an element $\sigma_t$ called state from some finite field $GF(2^m)$ and an updating rule. The rule of updating states is that multiplying $\sigma_t$ with $x^k$, $k$ is an integer, namely,

$$\sigma_{t+1} = x^k \cdot \sigma_t. \tag{1.1}$$

The finite fields used in here are $GF(2^m)$, $m = 128, 127, or\ 126$. In other word, the operation *shift* in LFSR now is replaced by multiplying $x^k$ in the field $GF(2^m)$.

The key sizes in DICING can be 128 bits or 256 bits, and the size of initial value may be taken as large as 256 bits, and the size of output of DICING is 128 bits.

In this paper the finite field $GF(2)$ is simply denoted as $\mathbb{F}$, and $\mathbb{F}[x]$ is the polynomial ring of unknown $x$ over the field $\mathbb{F}$. The symbols $\oplus$, $\otimes$ will represent the bitwise addition *XOR*, bitwise *and*, that is the operation & in $C$, and symbols $>>, <<, |$ and $\sim$ stand for the operations *right-shift, left-shift, concatenate* and *complement* respectively.

Suppose that $\zeta$ is a binary string, denoted by $\zeta[i]_{bit}$ and $\zeta[i,j]_{bit}$ the *i*-th bit and the segment from $i$-th bit to $j$-th bit respectively, and there are the similar expressions $\zeta[i]_{byte}$, $\zeta[i,j]_{byte}$ and $\zeta[i]_{word}$, $\zeta[i,j]_{word}$ measured in bytes and 32-bits words respectively, and if the meaning is explicit from the context, the low-index *bit*, *byte* and *word* will be omitted.

## 2. Construction

As general stream ciphers, the proposal cipher is also encrypt the plaintext and decrypt the ciphertext by adding bitwise a binary string called keystream, namely,

$$Ciphertext = Plain\,text \oplus Keystream \tag{2.1}$$

The keystream generator contains two main parts, a controller **H** and a combiner **C**. The controller **H** is made from two projectors $\Gamma_1$, $\Gamma_2$ and two counters $D'_t$, $D''_t$ which are also called dices. Denoted by $\alpha_t$ and $\beta_t$ the states of $\Gamma_1$ and $\Gamma_2$ in time $t$ respectively, which come from the finite fields $\mathbb{E}_1$ and $\mathbb{E}_2$ respectively, $\mathbb{E}_1 = \mathbb{F}[x]/p_1(x)$ and $\mathbb{E}_2 = \mathbb{F}[x]/p_2(x)$, $p_1(x)$ and $p_2(x)$ are the primitive polynomials with degree 127 and 126 respectively, which expression are given in the List 1. They satisfy the simple recurrence equations

$$\alpha_{i+1} = x^8 \cdot \alpha_i, \quad \beta_{i+1} = x^8 \cdot \beta_i, \quad i = 0, 1, 2, \ldots. \tag{2.2}$$

The dices $D'_t$ and $D''_t$ are two integers to record the last eight bits of $\alpha_t$ and $\beta_t$ respectively. The combiner **C** also contains two projectors $\Gamma_3$ and $\Gamma_4$, which are based on the two finite fields $\mathbb{E}_3$ and $\mathbb{E}_4$ respectively, $\mathbb{E}_3 = \mathbb{F}[x]/p_3(x)$, $\mathbb{E}_4 = \mathbb{F}[x]/p_4(x)$, $p_3(x)$ and $p_4(x)$ are primitive polynomials of degree 128 given in the List 1. Denoted by $\omega_t$ and $\tau_t$ the states of $\Gamma_3$ and $\Gamma_4$ in the time $t$ respectively, $\omega_t \in \mathbb{E}_3$ and $\tau_t \in \mathbb{E}_4$. Denoted by $D_t = (D'_t \oplus D''_t)$, $a = 1 + (D_t \,\&\, 15)$, $b = 1 + (D_t >> 4)$, they satisfy

$$\omega_{t+1} = x^a \cdot \omega_t, \quad \tau_{t+1} = x^b \cdot \tau_t. \tag{2.3}$$

Besides, we use two memorizes $u_t$ and $v_t$ to assemble $\omega_t$ and $\tau_t$ respectively,

$$u_t = u_{t-1} \oplus \omega_t, \quad v_t = v_{t-1} \oplus \tau_t, \quad for\ t > 0, \tag{2.4}$$

The initial values $\omega_0, \tau_0, u_0$ and $v_0$ will be specified in the later.

Suppose that $\mathbb{K}$ is a finite field $GF(2^8)$, $\mathbb{K} = \mathbb{F}[x]/p(x)$, $p(x)$ is an irreducible polynomial of degree eight, which expression is given in the List 1. We define S-box $S_0(x)$ as

$$S_0(x) = 5 \cdot (x \oplus 3)^{127}, \quad x \in \mathbb{K}. \tag{2.5}$$

We also adopt the representation $S_0(\zeta)$ for a bytes string $\zeta$ to represent that S-box $S_0$ substitute each byte of the string $\zeta$.

The startup includes two subprocesses *keysetup* and *ivsetup*, where the basic materials as the secret key and key-size will be input and the internal states will be initialized. Besides, in the *keysetup* we will make a key-defined the S-box $S(x)$ from $S_0(x)$ and a diffusion transformation $L$. The process is as following.

For a string $\rho$ of 8 bytes, we define an 8-bits vector $V_\rho$ and a $8 \times 8$ matrix $M_\rho$:

$$V_\rho[i] = \rho[8i+i]_{bit}, 0 \le i < 8, \qquad M_\rho = T_u \cdot J \cdot T_l. \tag{2.6}$$

where $T_u = (a_{i,j})_{8 \times 8}$ and $T_l = (b_{i,j})_{8 \times 8}$ are the upper-triangular matrix and the lower-triangular matrix respectively,

$$a_{i,j} = \begin{cases} \rho[8i+j]_{bit} & \text{if } i < j, \\ 1 & \text{if } i = j, \\ 0 & \text{if } i > j, \end{cases} \qquad b_{i,j} = \begin{cases} \rho[8i+j]_{bit} & \text{if } i > j, \\ 1 & \text{if } i = j, \\ 0 & \text{if } i < j, \end{cases} \tag{2.7}$$

and $J$ is a key-defined permutation matrix, for the simplicity, here take $J = 1$.

Suppose that $K$ is the secret key, let $\lambda = K[0,15]_{byte} \oplus K[16,31]_{byte}$, if $|K| = 256$, else $\lambda = K[0,15]_{byte}$, and $\lambda' = \lambda[0,7]_{byte}$, $\lambda'' = \lambda[8,15]_{byte}$, define two affine transformations on $\mathbb{K}$

$$A(x) = M_{\lambda'}(x), \qquad B(x) = M_{\lambda''}(x), \qquad x \in \mathbb{K}. \tag{2.8}$$

Denoted by $V_1 = V_{\lambda'} \oplus V_{\lambda''}$, and $V_2 = V_{\lambda'} \oplus ROTL8(V_{\lambda''}, 1)$, and then define a new S-box $S(x)$ and a transformation $L$ on $\mathbb{K}^4$,

$$S(x) = S_0(x \oplus V_2) \oplus V_1, \qquad L = \begin{pmatrix} A & B & A & A \oplus B \\ B & A & A \oplus B & A \\ A & A \oplus B & A & B \\ A \oplus B & A & B & A \end{pmatrix}. \tag{2.9}$$

Suppose that $\zeta$ is a string of $n$ bytes, if $n = 4k$ we also view it as a string of $k$ words, and

write $L(\zeta)$ to represent that $L$ takes on the each word of $\zeta$. Simply, we denote

$$Q(\zeta) = L \cdot S(\zeta). \tag{2.10}$$

In the *ivsetup*, the second step of the startup, the internal states will be initialized with the secret key and the initial value. In the generating keystream we will employ one mask of 16 bytes, which are denoted by $\eta$.

For a 32-bytes string $\zeta$ we define a bytes permutation $\phi$: $\zeta^\phi = \phi(\zeta)$, $\zeta^\phi[i] = \zeta[4i \bmod 31]$, for $0 \leq i < 31$, and $\zeta^\phi[31] = \zeta[31]$. Let $\hat{K} = K$ if $|K| = 256$ else $\hat{K} = K\,|\,(\sim K)$, denoted by $\check{K} = (\sim \hat{K}[16,31]_{byte})\,|\,(\sim \hat{K}[0,15]_{byte})$. We define the functions

$$F(\zeta) = Q(\phi(\zeta)), \quad G(\zeta) = F(F(F(\zeta) \oplus \hat{K}) \oplus \check{K}). \tag{2.11}$$

Suppose that $IV$ is the initial value of 32-bytes, $e$ is the base of natural logarithm and $c$ the integral part of $e \cdot 57!$, and $\xi_i$, $0 \leq i \leq 3$, are four 32-bytes strings defined as

$$\xi_0 = G(IV \oplus c), \quad \xi_i = G(\xi_{i-1} \oplus c), \quad i = 1, 2, 3. \tag{2.12}$$

The internal states are initialized respectively as following

$$\eta = \xi_0[0,15] \oplus \xi_0[16,31], \quad (u_0, v_0) = \xi_1, \quad (\alpha_0, \beta_0) = \xi_2', \quad (\omega_0, \tau_0) = \xi_3, \tag{2.13}$$

where $\xi_2' = \xi_2[0,126]\,|\,\xi_2[128,253]$, i.e. $\alpha_0 = \xi_2[0,126]$, $\beta_0 = \xi_2[128,253]$.

If $\xi_3 = 0$, the states $\omega_0$ and $\tau_0$ will be re-set as

$$(\omega_0, \tau_0) = \hat{K}. \tag{2.14}$$

*Note.* For a secret key, there is at most one $IV$ such that $\xi_3 = 0$.

After initializing, the process enters the recurrence part of generating keystream, each cycle includes two sub-processes of updating and combining. In the updating, all the states are updated from the time $t-1$ to the time $t$ as stated in (2.2) ~ (2.4). Suppose that $u$ and $v$ are two 16-bytes strings, which are also viewed as $4 \times 4$ matrices of bytes in the ordinary way. Denoted by $M^T$ the transposition of a matrix $M$, the combining function is defined as

$$C(u,v) = Q((Q(u) \oplus v)^T) \oplus \eta. \tag{2.15}$$

Denoted by $z_t$ the keystream in the time $t$, $(t > 0)$, then

$$z_t = C(u_t, v_t). \tag{2.16}$$

We have summarized the whole process in a sketch as Fig. 1.

List of the Primitive Polynomials used

| Polynomials | Expression |
|---|---|
| $p(x)$ | $x^8 + x^6 + x^5 + x + 1$ |
| $p_1(x)$ | $x^{127} + (x^{89} + x^{41} + 1)(x^3 + 1)$ |
| $p_2(x)$ | $x^{126} + (x^{83} + x^{35} + 1)(x^7 + 1)$ |
| $p_3(x)$ | $x^{128} + (x^{96} + x^{67} + x^{32} + 1)(x^3 + 1)$ |
| $p_4(x)$ | $x^{128} + (x^{96} + x^{64} + x^{37} + 1)(x^7 + x^5 + 1)$ |

List 1

The Sketch of Encryption Process

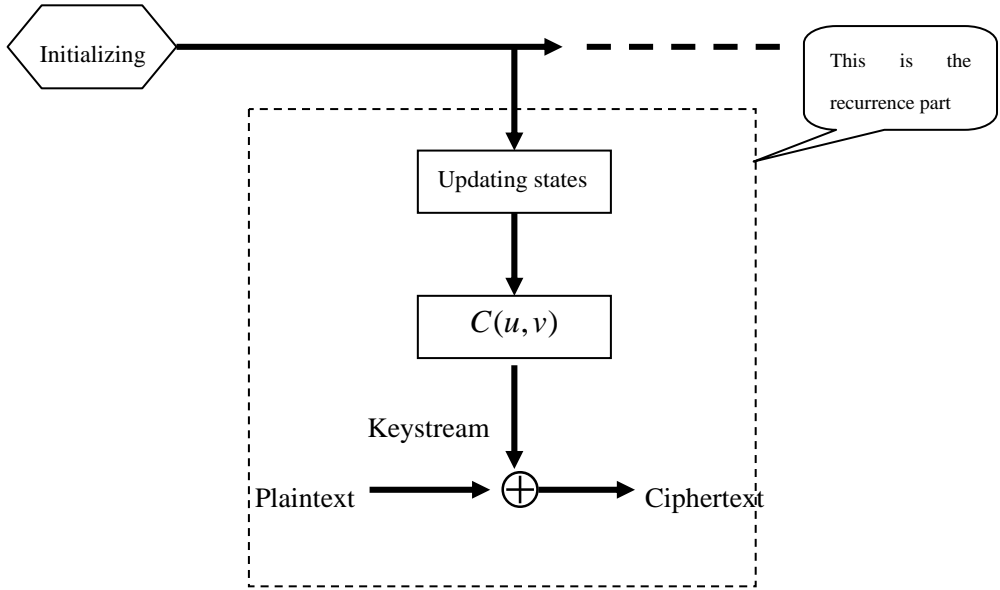

Fig.1

## 3. Security Analysis

In this section, we firstly show some results about the periods of the keystream of the stream cipher Dicing, and then give an investigation with respect to standard cryptanalytic attacks.

Denote $\pi(z_t)$ as the period of a sequence $z_t$.

**Proposition 1**

$$\pi(D_t) = (2^{126}-1)(2^{127}-1), \tag{3.1}$$

$$\pi(\omega_t) = \pi(\tau_t) = (2^{126}-1)(2^{127}-1)(2^{128}-1)/3 \tag{3.2}$$

Proof. Note that polynomials $p_1(x)$ and $p_2(x)$ are primitive, and the order of $x$ in the fields $\mathbb{E}_1$ and $\mathbb{E}_2$ are $2^{127}-1$ and $2^{126}-1$ respectively, hence

$$\pi(\alpha_t) = 2^{127}-1, \quad \pi(\beta_t) = 2^{126}-1, \tag{3.3}$$

and equation (3.1) is followed for $(2^{126}-1, 2^{127}-1) = 1$.

Write $\omega_i = \omega_{i-1} \cdot x^{k_i}$, and let $n = (2^{127}-1)(2^{126}-1)$, $m = \sum_{0<i\leq n} k_i$, it is easy to calculate that for each integer $c$, $1 < c \leq 16$, the occurrence times of $c$ in the sum above is $2^{249} - 2^{123} - 2^{122}$ and integer $1$ occurs one more times than the number. Thus,

$$m = (2^{249} - 2^{123} - 2^{122}) \cdot 136 + 1, \tag{3.4}$$

and

$$\omega_n = \omega_0 \cdot x^m = \omega_0 \cdot x^{1-2^{124}\cdot 5 \cdot 17} \tag{3.5}$$

In the field $\mathbb{E}_3$ or $\mathbb{E}_4$, the order of $x$ is equal to $2^{128}-1$, and $(2^{128}-1, 2^{124}\cdot 5\cdot 17 - 1) = 3$, the formula (3.2) is followed. □

Note that $u_t \oplus u_{t-1} = \omega_t$, $v_t \oplus v_{t-1} = \tau_t$, so we have

**Corollary 1**

$$\pi(u_t), \pi(v_t) \geq (2^{126}-1)(2^{127}-1)(2^{128}-1)/3. \tag{3.6}$$

In the next, we give a discussion for Dicing in respect to the resistance to the main known attacks.

We known that the correlation equations of the components in combining or filter functions should be known is required for the correlation attack, distinguishing attack and algebraic attack. In the proposed cipher, the updating of the components $\omega_t$ and $\tau_t$ are controlled by Pr.'s $\Gamma_1$ and $\Gamma_2$, and their correlations are not known for the adversaries, so these attacks will not be feasible.

As Dicing has a larger size of internal states, and their updating operations are multiplications of the finite field $GF(2^{128})$, and there are no the correspondences between some smaller isolated parts of the states and the keystream, hence there will be no flaws for the time/memory/data trade-off attacks and guess-and-determine attacks.

The initialization and combining functions also have protected DICING against the chosen-IV attacks, collision attacks and inversion attacks.

With our reference code, there are not found remarkable timing gaps for timing attack.

## 4. Implementation

In the platform of 32-bit Windows OS and AMD Athlon(tm) 64 x2 Dual Core processor 3600+, 2.00G Borland C++ 5.0, the performance of DICING is as following

Report of Performance

| Sub-processes | Time or Rate |
| --- | --- |
| Keysetup | 5200 *cycles* |
| IVsetup | 2280 *cycles* |
| Keystream Rate | 10 *cycles/byte* or 160 *cycles/block* |

List 2

## 5. The reduced versions and variants of DICING

The algorithm DICING presented above is a very conservative one, the users who wish to strive for a faster rate may adopt the reduced versions and variants. In the following are some such ones, and we think the reduced one will also have sufficient security.
1. Simply take the combining function

$$C(u,v) = Q(u) \oplus v, \qquad (5.1)$$

and omit the Pr. $\Gamma_2$.

2. In the case there are no attacks for conditional branching, the combining function may be taken as

$$C(u,v) = \begin{cases} Q(u) \oplus v & \text{if } \alpha[0] = 0, \\ Q(v) \oplus u & \text{if } \alpha[0] = 1. \end{cases} \quad (5.2)$$

3. Furthermore, omit both Pr. $\Gamma_1$ and Pr. $\Gamma_2$, and instead, let

$$dice = \omega[15] \oplus \tau[15]. \quad (5.3)$$

4. To obtain a larger period of keystream, the two Pr.'s, $\Gamma_3$ and $\Gamma_4$ may be replaced by a Pr. $\hat{\Gamma}$ from a finite field $GF(2^{256})$, i.e. $\mathbb{F}[x]/\hat{p}(x)$, $\hat{p}(x)$ is a primitive polynomial of degree 256. For 32-bit processors, $\hat{p}(x)$ may be taken as

$$\hat{p}(x) = x^{256} + (x^{224} + x^{192} + x^{161} + x^{128} + x^{96} + x^{67} + x^{32} + 1)(x^6 + 1). \quad (5.4)$$

The presented algorithm DICING was one of candidates for eSTREAM [1], [2], but with a little difference from [2] in (2.8) and (2.9), here the vector $V_1$ has been moved into S-box $S(x)$ from the transformation $A(x)$.